\documentclass[number,preprint,3p]{elsarticle}

\usepackage{setspace}
\usepackage{nicefrac}
\usepackage{color}
\usepackage{graphicx}
\usepackage{subcaption}
\usepackage{algorithm}
\usepackage{algpseudocode}
\usepackage{bm}
\usepackage[colorlinks]{hyperref}
\usepackage{amssymb}
\usepackage{amsthm}
\usepackage{amsmath}
\usepackage{mathtools}
\usepackage{dsfont}
\usepackage{booktabs}
\usepackage{threeparttable}
\usepackage{tabularx}
\usepackage{multirow}
\usepackage{array}
\usepackage{xurl}
\usepackage{epstopdf}
\usepackage{float}
\usepackage{tcolorbox}
\usepackage{tablefootnote}
\usepackage[perpage]{footmisc}
\usepackage{bbm}
\usepackage{lineno}

\newcommand{\Prob}[1]{\mathbb{P}\left(#1\right)}
\newcommand{\E}[1]{\mathbb{E}\left[#1\right]}

\newcommand{\vect}[1]{\boldsymbol{#1}}




\makeatletter
\gdef\urlauthor#1#2{\g@addto@macro\@elsuads{\let\corref\@gobble%
     \def\@@tmp{#1}\raggedright\eadsep
     {\ttfamily\url{\expandafter\strip@prefix\meaning\@@tmp}}\space(#2)%
     \def\eadsep{\unskip,\space}}%
}
\gdef\emailauthor#1#2{\stepcounter{ead}%
     \g@addto@macro\@elseads{\raggedright%
      \let\corref\@gobble\def\@@tmp{#1}%
      \eadsep{\ttfamily\href{mailto:\expandafter\strip@prefix\meaning\@@tmp}{\expandafter\strip@prefix\meaning\@@tmp}}
      (#2)\def\eadsep{\unskip,\space}}%
}
\makeatother

\biboptions{numbers,sort&compress,square}

\journal{arXiv}

\begin{document}
\begin{frontmatter}

\renewcommand{\thefootnote}{\fnsymbol{footnote}}

\title{A Statistical Physics View of the S\&P 500: Pairwise Interactions and Time-Varying Dynamics}

\author[1]{Sebin Oh}
\author[1,2,3]{Marta C. Gonz\'{a}lez\corref{cor1}}
\ead{martag@berkeley.edu}
\author[1]{Ziqi Wang\corref{cor1}}
\ead{ziqiwang@berkeley.edu}
\cortext[cor1]{Corresponding authors}
\address[1]{Department of Civil and Environmental Engineering, University of California, Berkeley, United States}
\address[2]{Department of City and Regional Planning, University of California, Berkeley, United States}
\address[3]{Lawrence Berkeley National Laboratory, Berkeley, United States}

\begin{abstract}
We analyze a fixed panel of S\&P 500 stocks from 1996 to 2026 using complementary static and kinetic Ising models applied to daily binary open-to-close movements. The static pairwise model provides a long-run maximum-entropy summary of low-order dependence and reveals a sectorally organized interaction network with modest small-world structure and within-sector couplings about 2.8 times stronger than between-sector couplings, with especially coherent real estate and energy sectors. The kinetic model incorporates smooth time-varying external fields, self-memory, and directed lagged couplings to describe next-day dynamics. It reveals slow field-regime shifts around three major market-wide perturbations---the dot-com bust, the global financial crisis, and the COVID-19 episode. Self-memory is generally weak, and the directed coupling structure is much less sector-concentrated and more asymmetric than the static network, while still reproducing the broad evolution of aggregate market movement. Taken together, the two complementary models characterize both persistent market organization and short-horizon cross-stock dynamics, providing a compact statistical physics view of interaction structure and time-varying behavior in the S\&P 500.
\end{abstract}


\end{frontmatter}

\renewcommand{\thefootnote}{\fnsymbol{footnote}}

\section{Introduction}\label{sec:intro}\noindent
Financial markets are high-dimensional systems in which many assets interact through common information, sectoral linkages, and institutional connections~\cite{mantegna_hierarchical_1999,onnela_dynamics_2003,tumminello_correlation_2010,bardoscia_physics_2021,Tang_et_al._2018}. These interdependencies make the system inherently complex and can produce collective behavior such as synchronization and phase transitions~\cite{kim_agent-based_2008,zhang_financial_2023}. This complexity has motivated a wide range of efforts to characterize market-wide structure and collective behavior.

A major strand of research has analyzed financial markets through complex-network methods. Correlation matrices, minimum spanning trees, planar maximally filtered graphs, and related dependency networks have been used to uncover clustering, hierarchy, centrality, and topological changes during crises~\cite{mantegna_hierarchical_1999,onnela_dynamics_2003,tumminello_correlation_2010,kenett_dominating_2010,Tang_et_al._2018,rakib_structure_2021}. These approaches provide useful structural descriptions, but the inferred networks depend on choices such as the dependence measure, filtering procedure, and temporal aggregation. They also do not, by themselves, specify a probabilistic model for joint market states. Recent work has further emphasized that financial networks should be studied as dynamical systems rather than purely static objects~\cite{kobayashi_social_2018}, and that changes in stock-network organization can be informative about market instability and volatility~\cite{gorduza_understanding_2025}.

Maximum entropy modeling offers a complementary probabilistic framework by inferring the least-structured distribution consistent with selected empirical constraints~\cite{jaynes_information_1957,nguyen_inverse_2017}. For binary variables, constraining first- and second-order moments yields the pairwise Ising model. This framework has been applied widely in diverse fields, including machine learning, ecology, neuroscience, and natural hazard resilience~\cite{aurell_inverse_2012,nguyen_inverse_2017,schneidman_weak_2006,oh_longrange_2024,oh_phase_2026}. Within finance, maximum-entropy studies have focused mainly on static pairwise dependence and collective market states~\cite{bury_market_2013,bury_statistical_2013,zhang_financial_2023}; comparatively less attention has been given to combining long-run interaction structure and time-varying directed dynamics within a unified Ising-based framework.

This study analyzes a fixed 30-year panel of stock data from firms in the S\&P 500, a standard benchmark for large-cap U.S. equities, using complementary static and kinetic Ising models. The static pairwise Ising model provides a least-biased long-run summary of low-order dependence and the associated interaction network over the full sample. The kinetic Ising model complements it by exploiting the temporal ordering of the data through time-varying external fields, self-memory, and directed cross-stock interactions. This combined perspective is useful because a full-sample static fit captures the persistent low-order structure, while a dynamic model offers a statistical description of the transient, next-day market dynamics. Estimating both models on the same set of firms over the same sample period permits a direct comparison between long-run interaction structure and evolving short-run directional dynamics. We therefore examine both how the market is organized as an interacting system over the long run and how influence propagates across that system from one trading day to the next. By integrating these two perspectives within a unified maximum-entropy framework, our approach extends prior work that often isolates static network topology from dynamic propagation, yielding a more comprehensive and probabilistically grounded characterization of market complexity.

This paper is organized as follows. Section~\ref{sec:data} describes the data construction and preprocessing steps. Section~\ref{sec:ising} introduces the static and kinetic Ising models and the associated estimation procedures. Section~\ref{sec:static_net} presents the long-run interaction network inferred from the static model. Section~\ref{sec:kinetic_dyn} studies the time-varying market dynamics captured by the kinetic model. Section~\ref{sec:conclusion} concludes.

\section{Data and preprocessing}\label{sec:data}\noindent
We construct a daily binary panel for a fixed set of S\&P 500 firms spanning January 1, 1996 to January 1, 2026. Company symbols, sector labels, and S\&P 500 inclusion dates are collected from the publicly available S\&P 500 constituent table. Sector labels follow the Global Industry Classification Standard (GICS)~\cite{gics_methodology}, which categorizes companies into the 11 primary sectors summarized in Table~\ref{tab:gics_sectors}. Daily open and close prices are downloaded from Yahoo Finance using the Python package \texttt{yfinance}~\cite{ranaroussi_yfinance}.

\begin{table}[H]
    \centering
    \caption{\textbf{The 11 Global Industry Classification Standard (GICS) sectors and their abbreviations used in this study.} These standard sector definitions are used to categorize the 306 S\&P 500 firms retained in the fixed panel.}
    \label{tab:gics_sectors}
    \small
    \begin{tabular}{@{}ll@{}}
    \toprule
    \textbf{Sector} & \textbf{Abbreviation} \\
    \midrule
    Communication Services & Comm \\
    Consumer Discretionary & ConsDisc \\
    Consumer Staples & Staples \\
    Energy & Energy \\
    Financials & Fin \\
    Health Care & Health \\
    Industrials & Ind \\
    Information Technology & IT \\
    Materials & Mat \\
    Real Estate & RE \\
    Utilities & Util \\
    \bottomrule
    \end{tabular}
\end{table}

For each stock $i$ on each trading day $t$, we binarize the daily open-to-close movement into a spin variable $s_i(t)\in\{-1,+1\}$, assigning $s_i(t)=+1$ when the closing price exceeds the opening price and $s_i(t)=-1$ otherwise. Observations with missing open or close prices are discarded.

\begin{figure}[H]
    \centering
    \includegraphics[width=1.0\linewidth]{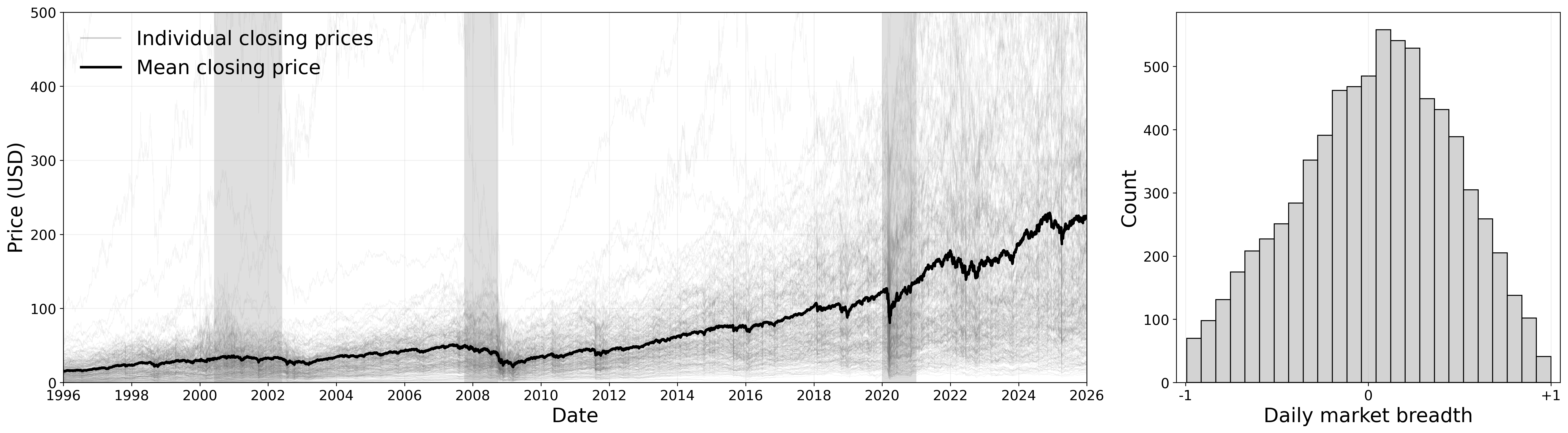}
    \caption{\textbf{Raw closing-price panel and empirical distribution of daily market breadth.} Left: gray lines show the daily closing prices of the 306 firms, and the black line shows the unweighted cross-sectional mean closing price. Shaded gray bands mark the dot-com collapse (June 1, 2000--June 1, 2002), the global financial crisis (October 1, 2007--October 1, 2008), and the COVID-19 period (January 1, 2020--January 1, 2021). A shorter drawdown is also visible around April 2025, coinciding with tariff-related market volatility. Right: histogram of daily market breadth across the full sample, computed as the cross-sectional mean of the binarized stock states, $\frac{1}{N}\sum_{i=1}^{N}s_i(t)$ where $N=306$ denotes the number of stocks in the panel.}
    \label{fig:sp500_close_with_hist}
\end{figure}

We retain only firms with complete observations over the entire 30-year period. This yields a fixed panel of 306 companies observed over 7,550 trading days, corresponding to a data matrix of size \((7550,306)\). The same panel is used for both the static and kinetic Ising models so that the two results are comparable. Figure~\ref{fig:sp500_close_with_hist} shows the corresponding raw closing-price panel before binarization together with the empirical distribution of daily market breadth after binarization, where daily market breadth is defined as the cross-sectional mean of the binarized stock states. The raw series display large cross-sectional differences in prices, persistent long-run drift, and common declines during major market-wide perturbations. These features motivate the binary encoding adopted here, which removes non-comparable level effects across firms and centers the analysis on directional movement rather than on raw price magnitudes.

\section{Maximum entropy models for stock movements}\label{sec:ising}\noindent
As a maximum entropy model for binary variables under low-order moment constraint, the Ising model yields a static model that matches first- and second-order moments and a kinetic model that matches one-step transition statistics. This section introduces both formulations and their estimation; \ref{app:maxent} provides the corresponding derivations using the principle of maximum entropy.

\subsection{Static Ising model}\label{sec:ising_sta}\noindent
For binary stock movements, let $s_i\in\{-1,+1\}$ denote the movement of stock $i$, with $+1$ and $-1$ representing upward and downward moves, respectively. The static Ising model assigns to a configuration $\vect{s}$ the Hamiltonian
\begin{equation}
    \mathcal{H}(\vect{s})=-\sum_{i=1}^N h_i s_i - \sum_{i<j} J_{ij}s_i s_j,
\end{equation}
where $N$ is the number of stocks, $h_i$ represents a local field acting on spin $i$, and $J_{ij}$ represents the pairwise coupling between spins $i$ and $j$. In the financial context, \(h_i\) reflects the intrinsic tendency of stock \(i\) to move upward or downward, while \(J_{ij}\) captures pairwise dependence between stocks arising from shared exposure to economic factors or correlated investor behavior. A positive $J_{ij}$ favors alignment, whereas a negative $J_{ij}$ favors opposite movement. In the static Ising model, the coupling matrix $\vect{J}=[J_{ij}]$ is symmetric by construction and can be interpreted as an undirected effective interaction network.

The probability of observing $\vect{s}$ is then given by the Boltzmann distribution $p(\vect s) \propto \exp(-\beta \mathcal{H}(\vect{s}))$, where $\beta$ represents the inverse temperature. In data-analytic applications~\cite{nguyen_inverse_2017}, $\beta$ is usually absorbed into the parameters, giving
\begin{equation}\label{eq:static_ising_prob}
   p(\vect s)=\frac{1}{Z}\exp\biggl(\sum_{i=1}^N h_i s_i+\sum_{i<j}J_{ij}s_i s_j\biggr),
\end{equation}
where $Z$ is the normalization constant (partition function).

\subsection{Kinetic Ising model}\label{sec:ising_dyn}\noindent
The kinetic Ising model~\cite{mezard_exact_2011,roudi_dynamical_2011, campajola_modelling_2022} specifies the conditional probability of the next market state given the current state. We adopt a synchronous update rule, meaning that all stock movements are updated simultaneously at each time step. This choice is natural for stock market data because all stocks share the same daily opening and closing times. Under this formulation, the conditional probability of the next state $\vect{s}(t+1)$ given the current state $\vect{s}(t)$ is
\begin{equation}\label{eq:kinetic_ising_prob}
    p\left(\vect{s}(t+1)\mid\vect{s}(t)\right)
    =
    \prod_{i=1}^N
    \frac{\exp\!\left(s_i(t+1)\,\theta_i(t)\right)}
    {2\cosh\!\left(\theta_i(t)\right)},
\end{equation}
where the effective local field is
\begin{equation}
    \theta_i(t)=h_i(t)+a_i s_i(t)+\sum_{j\neq i} J_{ij}s_j(t).
\end{equation}
Here $J_{ij}$ measures how the state of stock $j$ at time $t$ affects the tendency of stock $i$ at time $t+1$, $h_i(t)$ represents time-varying external influences, and $a_i$ is a self-memory coefficient that allows the next-day movement of stock $i$ to depend on its past state at time $t$. Unlike the static model, the kinetic coupling matrix $\vect{J}$ is not necessarily symmetric. A positive $a_i$ indicates short-term persistence, whereas a negative $a_i$ indicates short-term reversal. Equation~\eqref{eq:kinetic_ising_prob} implies
\begin{equation}\label{eq:kinetic_prediction}
    \E{s_i(t+1)\mid\vect{s}(t)}
    =
    \tanh\!\left(\theta_i(t)\right).
\end{equation}

From a parameter-estimation perspective, allowing $h_i(t)$ to vary independently at each time step would substantially increase the number of parameters to be estimated, thereby hindering both reliable estimation and interpretability. We therefore represent $h_i(t)$ using a basis expansion,
\begin{equation}
    h_i(t)=\sum_{m=1}^{M}\phi_m(t)\gamma_{im},
\end{equation}
where $\phi_m(t)$ is the $m$th temporal basis function. Specifically, we adopt a piecewise-linear hat basis,
\begin{equation}
    \phi_m(t)=\max\!\left\{1-\frac{|\tau(t)-c_m|}{\Delta},\,0\right\},
    \qquad
    \Delta=\frac{1}{M-1},
\end{equation}
where $\tau(t)=\frac{t-1}{T-2}$ for $t=1,\ldots,T-1$ is the normalized time index, and $c_m=\frac{m-1}{M-1}$ for $m=1,\ldots,M$ denotes the equally spaced center of the $m$th basis function. Thus, each basis function has local support and overlaps linearly only with its neighboring basis functions. This yields
\begin{equation}
    \theta_i(t)
    =
    \sum_{m=1}^{M}\phi_m(t)\gamma_{im}
    + a_i s_i(t)
    + \sum_{j\neq i} J_{ij}s_j(t),
\end{equation}
where $\gamma_{im}$ is the coefficient of the $m$th basis function for stock $i$. We use $M=30$ basis functions over the 30-year observation period, so this representation is sufficiently flexible to capture year-scale variation in external effects. Under this specification, the effective local field consists of three components: the cross-stock interaction term $\sum_{j\neq i} J_{ij}s_j(t)$, the self-memory term $a_i s_i(t)$, and the time-varying external field $h_i(t)$.

This formulation of the kinetic Ising model, which may also be viewed as a discrete-time Markov chain on the finite state space $\{-1,+1\}^N$, is often used as a nonequilibrium model of binary dynamics~\cite{levin2017markov,aguilera_unifying_2021}. Under special choices of parameters and update rules the chain may satisfy \emph{detailed balance} and admit the static Ising distribution as a stationary distribution.

\subsection{Parameter estimation}\label{sec:ising_est}\noindent
For the full 30-year sample, we estimate the static Ising parameters $\vect{h}$ and $\vect{J}$ by maximum likelihood~\cite{aurell_inverse_2012,nguyen_inverse_2017}. Let $\mathcal{D}=\{\vect{s}^{(1)},\vect{s}^{(2)},\ldots,\vect{s}^{(n_{\mathrm{obs}})}\}$ denote the set of $n_{\mathrm{obs}}$ daily stock movement configurations, where each $\vect{s}^{(m)}\in\{-1,+1\}^N$. The average log-likelihood is
\begin{equation}
    \mathcal{L}_{\mathrm{static}}(\vect{h},\vect{J}\mid\mathcal{D})
    =
    \sum_{i=1}^{N} h_i \mathcal{M}_{1,i}^\mathcal{D}
    + \sum_{i<j} J_{ij} \mathcal{M}_{2,ij}^\mathcal{D}
    - \log Z(\vect{h},\vect{J}),
\end{equation}
where
\(
    \mathcal{M}_{1,i}^\mathcal{D}
    =
    \frac{1}{n_{\mathrm{obs}}}\sum_{m=1}^{n_{\mathrm{obs}}}s_i^{(m)}
\)
and
\(
    \mathcal{M}_{2,ij}^\mathcal{D}
    =
    \frac{1}{n_{\mathrm{obs}}}\sum_{m=1}^{n_{\mathrm{obs}}}s_i^{(m)}s_j^{(m)}
\)
are the empirical first- and second-order moments. The corresponding gradient components are
\begin{equation}
    \frac{\partial \mathcal{L}_{\mathrm{static}}}{\partial h_i}
    =
    \mathcal{M}_{1,i}^\mathcal{D}-\mathcal{M}_{1,i},
    \qquad
    \frac{\partial \mathcal{L}_{\mathrm{static}}}{\partial J_{ij}}
    =
    \mathcal{M}_{2,ij}^\mathcal{D}-\mathcal{M}_{2,ij},
\end{equation}
so maximum likelihood amounts to matching the empirical moments to the model counterparts calculated using Eq.~\eqref{eq:static_ising_prob}. Because exact evaluation of $Z$ and of the model moments $\mathcal{M}_{1,i}$ and $\mathcal{M}_{2,ij}$ is infeasible for a system of this size, we use Monte Carlo maximum likelihood, with Gibbs sampling to approximate the model moments during gradient-based optimization~\cite{Maoz2017_191625}.

For the kinetic Ising model, the parameters are estimated from the same time-ordered data $\mathcal{D}=\{\vect{s}(1),\vect{s}(2),\ldots,\vect{s}(T)\}$, where $T=n_{\mathrm{obs}}$, by penalized conditional maximum likelihood, with average conditional log-likelihood
\begin{equation}
    \mathcal{L}_{\mathrm{kinetic}}
    =
    \frac{1}{T-1}\sum_{t=1}^{T-1}\sum_{i=1}^{N}
    \left[
        s_i(t+1)\theta_i(t)-\log\!\left(2\cosh\!\left(\theta_i(t)\right)\right)
    \right].
\end{equation}
Because the synchronous conditional probability factorizes over nodes (Eq.~\eqref{eq:kinetic_ising_prob}), the conditional log-likelihood decomposes into a sum of stock-specific terms, and the parameters associated with each stock can be estimated separately. For each stock $i$, the corresponding gradient components are
\begin{equation}
    \frac{\partial \mathcal{L}_{\mathrm{kinetic}}}{\partial \gamma_{im}}
    =
    \frac{1}{T-1}\sum_{t=1}^{T-1}
    \phi_m(t)\left[s_i(t+1)-\tanh\!\left(\theta_i(t)\right)\right],
\end{equation}
\begin{equation}
    \frac{\partial \mathcal{L}_{\mathrm{kinetic}}}{\partial a_i}
    =
    \frac{1}{T-1}\sum_{t=1}^{T-1}
    s_i(t)\left[s_i(t+1)-\tanh\!\left(\theta_i(t)\right)\right],
\end{equation}
and
\begin{equation}
    \frac{\partial \mathcal{L}_{\mathrm{kinetic}}}{\partial J_{ij}}
    =
    \frac{1}{T-1}\sum_{t=1}^{T-1}
    s_j(t)\left[s_i(t+1)-\tanh\!\left(\theta_i(t)\right)\right],
    \qquad j\neq i.
\end{equation}
These gradient expressions show that, for the unpenalized conditional likelihood, estimation amounts to comparing the observed next-step movement $s_i(t+1)$ with its conditional expectation $\tanh\!\left(\theta_i(t)\right)$. To stabilize the fit, we add $\ell_2$ penalties to $\gamma_{im}$, $a_i$, and $J_{ij}$, together with a second-difference penalty on the basis coefficients $\gamma_{im}$ across $m$ to encourage smooth variation in the external field. Unlike in the static case, this conditional likelihood is tractable and does not require evaluation of a global partition function.

\section{Interaction structure in the static Ising model}\label{sec:static_net}\noindent
The fitted static Ising model yields a signed, weighted interaction network over the S\&P 500 stocks. As shown in Appendix Fig.~\ref{fig_apdx:validation_static}, the model closely reproduces the empirical first- and second-order moments, indicating that the inferred external fields $h_i$ and couplings $J_{ij}$ capture the dominant low-order dependence structure of the pooled 30-year panel. For the filtered interaction network, the small-world coefficient is $\sigma = \frac{C/C_{\mathrm{rand}}}{L/L_{\mathrm{rand}}}=1.424$, where $C$ and $L$ denote the observed average clustering coefficient and average shortest-path length, respectively, and $C_{\mathrm{rand}}$ and $L_{\mathrm{rand}}$ denote the corresponding edge-matched random-graph benchmarks. Thus, $\sigma>1$ suggests that the network shows a modest small-world tendency: the network has higher clustering than comparable random graphs while maintaining a similarly short path length. Detailed diagnostics are reported in Table~\ref{tab_apdx:static_network_diagnostics} in the Appendix, with benchmark comparison provided in Appendix Fig.~\ref{fig_apdx:benchmark_comparison}. The network also exhibits positive sector assortativity of $0.259$, reflecting a tendency for stocks to connect more densely with other stocks in the same sector and indicating that sector membership remains an important organizing principle. The discussion below focuses on the structural patterns revealed by the inferred fields and couplings.

\begin{figure}[H]
    \centering
    \includegraphics[width=1.0\linewidth]{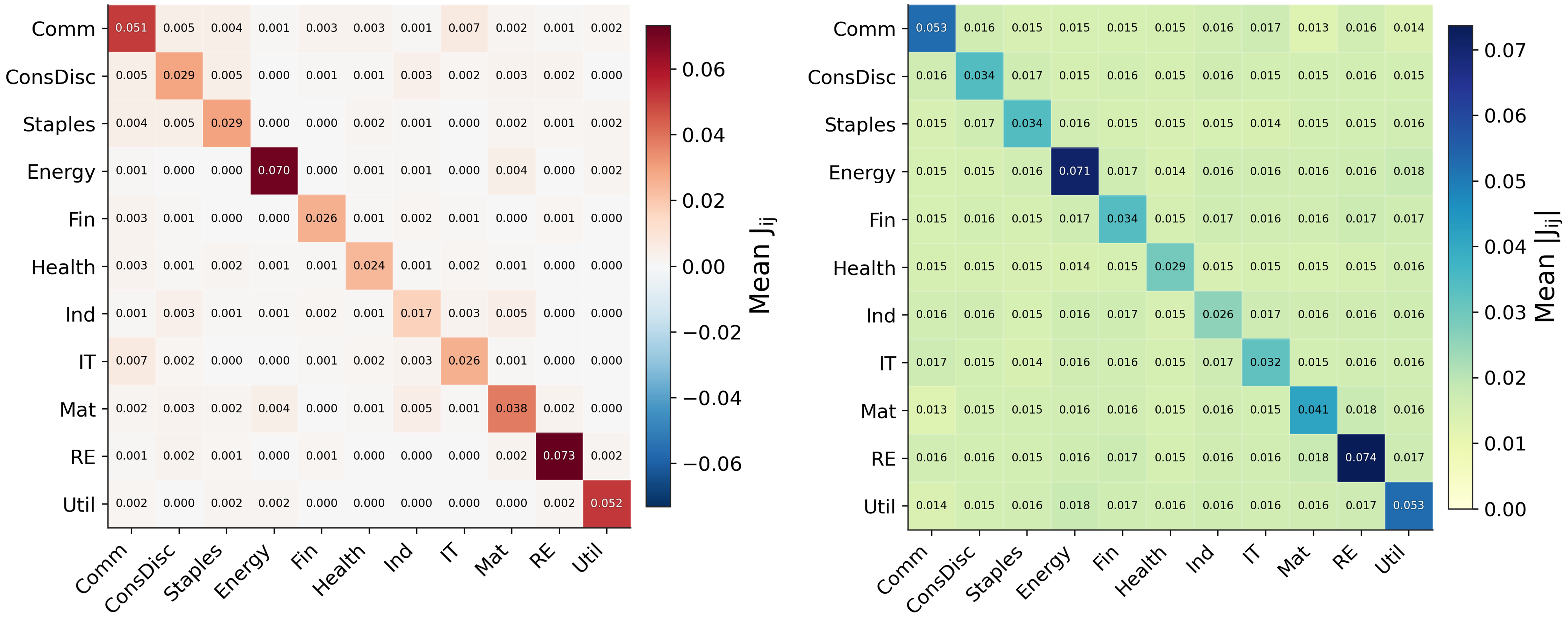}
    \caption{\textbf{Sector-level summaries of the static Ising couplings.} Left: mean signed coupling $J_{ij}$ for each sector pair. Right: mean absolute coupling $|J_{ij}|$ for each sector pair. Diagonal entries denote within-sector averages, whereas off-diagonal entries denote between-sector averages. Strong diagonal dominance indicates systematically stronger within-sector interactions, with especially large within-sector magnitudes in real estate and energy. The signed averages are weakly positive for most sector pairs.}
    \label{fig:sector_heatmaps}
\end{figure}

\begin{figure}[H]
    \centering
    \includegraphics[width=0.5\linewidth]{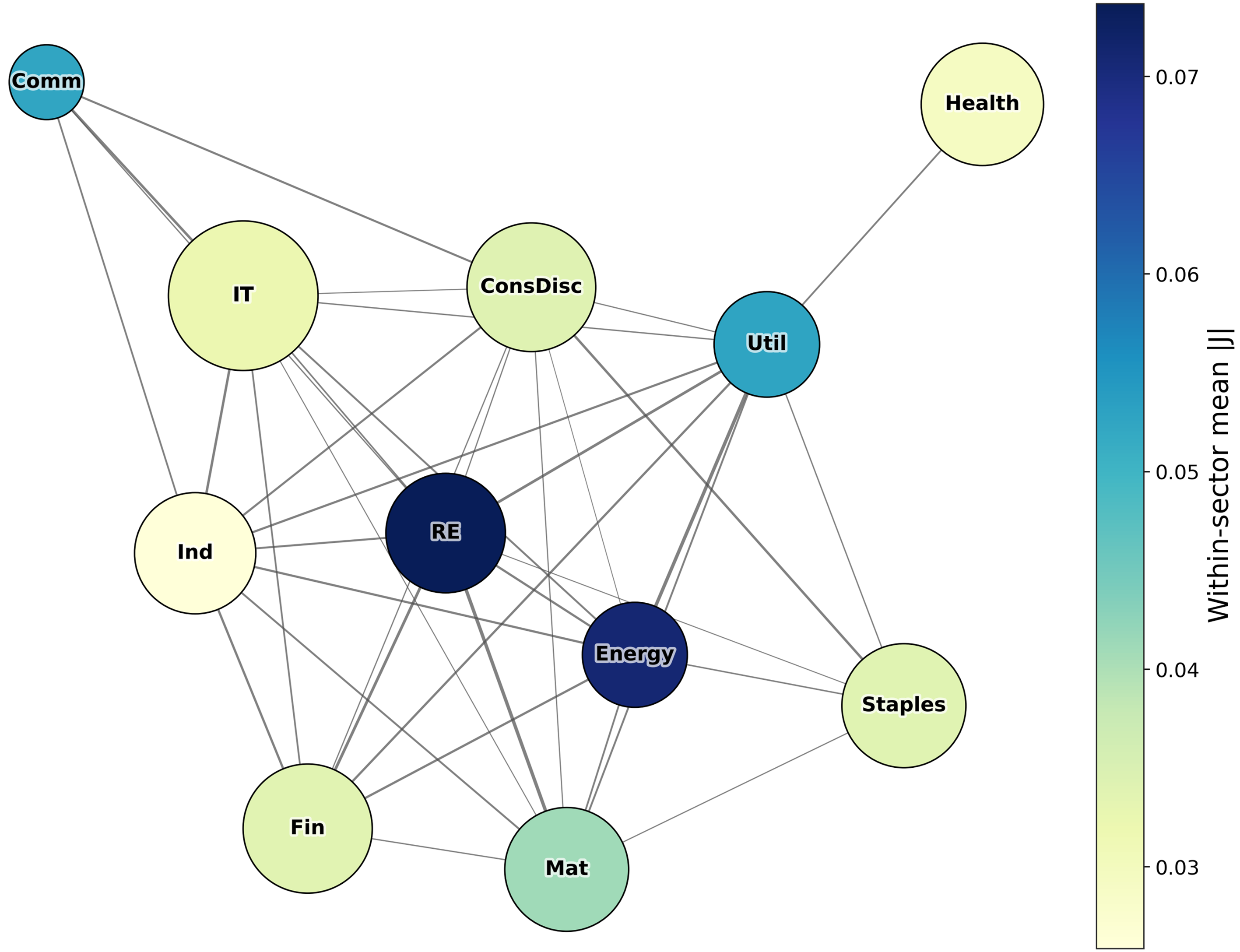}
    \caption{\textbf{Sector interaction network.} Node size is proportional to the sector mean $|h_i|$, node color indicates the mean within-sector $|J_{ij}|$, and edge width is proportional to the mean absolute inter-sector coupling. To avoid an overly dense display, only inter-sector links above the 30th percentile of cross-sector mean absolute couplings are shown, corresponding approximately to the strongest 70\% of cross-sector links. The retained cross-sector edges have broadly comparable widths, indicating that no single sector pair dominates the inter-sector interaction pattern. Health care and communication services appear relatively selective, with fewer retained links than most other sectors.}
    \label{fig:sector_network}
\end{figure}

Figure~\ref{fig:sector_heatmaps} summarizes the coupling structure in matrix form. Diagonal entries dominate both the signed and absolute sector-level matrices, and the mean within-sector interaction magnitude is about 2.8 times the between-sector average. Real estate and energy stand out as the most coherent sectors, whereas most between-sector signed averages remain small and weakly positive. Figure~\ref{fig:sector_network} provides a complementary sector-level view in which node size is proportional to the sector mean $|h_i|$, node color reflects the mean within-sector $|J_{ij}|$, and edge width represents the mean absolute inter-sector coupling. Information technology has the largest field magnitude, with health care also prominent on this dimension, while real estate and energy remain most prominent in terms of within-sector coupling. The retained cross-sector links have broadly similar widths, consistent with the relatively homogeneous off-diagonal values in the absolute sector matrix, so no single sector pair dominates the cross-sector interaction pattern. Health care and communication services appear more selectively connected. Taken together, the sector-level summaries depict a market organized around strong within-sector cohesion while remaining tied together through broadly distributed cross-sector interactions.

\begin{figure}[H]
    \centering
    \includegraphics[width=0.9\linewidth]{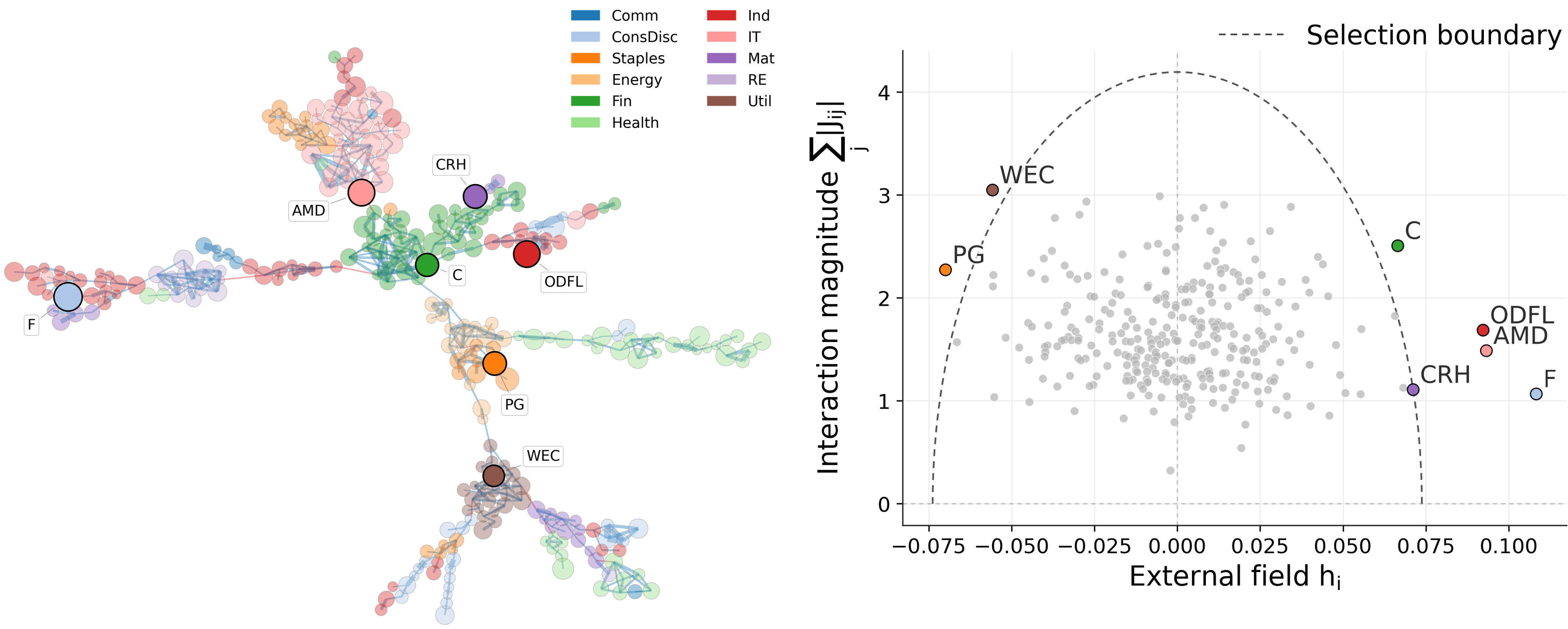}
    \caption{\textbf{Company-level backbone structure and stocks with large field and interaction strength in the full 30-year network.} Left: stock-level backbone of the inferred interaction network. The full-edge interaction network is first filtered by retaining the edges with the top 10\% $\lvert J_{ij} \rvert$ values, corresponding to the cutoff \(\tau_J=0.03534\), yielding 4{,}667 retained edges in the filtered network. For visualization, the backbone is then constructed from this filtered network using a maximum-spanning-tree scaffold based on $\lvert J_{ij} \rvert$, with the strongest remaining non-tree edges added to yield 467 edges in total, approximately 10\% of the filtered-network edges. Nodes are colored by sector and sized by $|h_i|$, while edge width is proportional to $|J_{ij}|$. Right: scatter plot of external field $h_i$ versus aggregate absolute coupling $\sum_j |J_{ij}|$, where each point represents a stock. Gray points show all stocks, whereas colored labeled points denote the highlighted stocks with large field and interaction strength also shown in the backbone: WEC (WEC Energy Group), PG (Procter \& Gamble), C (Citigroup), AMD (Advanced Micro Devices), ODFL (Old Dominion Freight Line), F (Ford Motor Company), and CRH (CRH plc). After rescaling $|h_i|$ and $\sum_j |J_{ij}|$ by their respective 90th-percentile values, these stocks are identified as the top 2\% with the largest normalized radial distances. The dashed curve in the right panel shows the corresponding selection boundary.}
    \label{fig:stock_network_leading_candidates}
\end{figure}

The stock-level backbone in Fig.~\ref{fig:stock_network_leading_candidates} adds finer resolution to the sector-level picture. Since the full interaction matrix is dense and dominated by near-zero entries (Appendix Fig.~\ref{fig_apdx:static_ising_params_histograms}), we first construct a filtered network by retaining the top decile of pairwise coupling magnitudes $\lvert J_{ij} \rvert$, equivalently thresholding with $\lvert J_{ij} \rvert=0.03534$ (Table~\ref{tab_apdx:static_network_diagnostics}). The left panel compresses this filtered network into a sparse backbone that preserves connectivity and the strongest pairwise relations, revealing compact clusters that largely follow sector boundaries together with a smaller set of cross-sector connections. The right panel places all stocks in the $(h_i,\sum_j |J_{ij}|)$ plane, where distance from the central cloud reflects the combined magnitude of the external field and aggregate coupling strength. A large value of $\sum_j |J_{ij}|$ indicates that a stock is strongly tied to the rest of the network, so stocks with both large $|h_i|$ and large aggregate coupling occupy especially prominent positions in the inferred market structure. Most stocks remain near the center, whereas WEC Energy Group (WEC), Procter \& Gamble (PG), Citigroup (C), Advanced Micro Devices (AMD), Old Dominion Freight Line (ODFL), Ford Motor Company (F), and CRH plc (CRH) lie outside the selection boundary. WEC and PG stand out mainly through unusually large aggregate coupling strength despite negative fields; C, AMD, and ODFL combine strong coupling with large positive fields; CRH is elevated on both dimensions more moderately; and F is distinguished primarily by an extreme positive field. These firms therefore emerge as the most prominent stocks in the inferred 30-year interaction network. Notably, the highlighted firms are distributed across different parts of the backbone rather than concentrated in a single cluster, indicating that prominence in the inferred interaction structure is not confined to one sector.

Overall, the static Ising network portrays the S\&P 500 as sectorally organized yet not sectorally fragmented, with modest small-world properties. The strongest dependencies concentrate within sectors, yet cross-sector couplings keep those sectors embedded in a single market-wide structure. A limited set of firms also stands out through the combination of large field magnitude and strong aggregate coupling.

The static fit serves as a long-run low-order interaction surrogate for the market’s backbone structure. By construction, however, it compresses the full 30-year sample to a single equilibrium description up to low-order moments and therefore cannot capture temporal nonstationarity or regime mixing, as reflected in the mismatch between the empirical and model-implied distributions of daily market breadth (Appendix Fig.~\ref{fig_apdx:validation_static_magnetization}). This limitation motivates the kinetic Ising analysis, which uses the temporal ordering of the data to incorporate slowly varying background conditions, self-memory, and directed lagged interactions.

\section{Market dynamics in the kinetic Ising model}\label{sec:kinetic_dyn}\noindent
The kinetic Ising model uses the time ordering of the data to describe next-day market movement through a slowly varying external field $h_i(t)$, a self-memory term $a_i$, and directed interactions $J_{ij}$ that transmit influence across stocks. The field, represented through a low-dimensional basis expansion, captures market-wide low-frequency perturbations rather than day-to-day shocks. The inferred dynamics therefore separate naturally into two layers: a low-frequency regime component carried by the external field and a high-frequency day-to-day component carried mainly by the directed interaction network. Appendix Fig.~\ref{fig_apdx:validation_kinetic} shows that the fitted model is well calibrated and closely reproduces the stock-level means and lagged cross-stock correlations. More detailed fit diagnostics are reported in Appendix Table~\ref{tab_apdx:kinetic_diagnostics}. The discussion below therefore focuses on the broader dynamical picture.

\begin{figure}[H]
    \centering
    \includegraphics[width=1.0\linewidth]{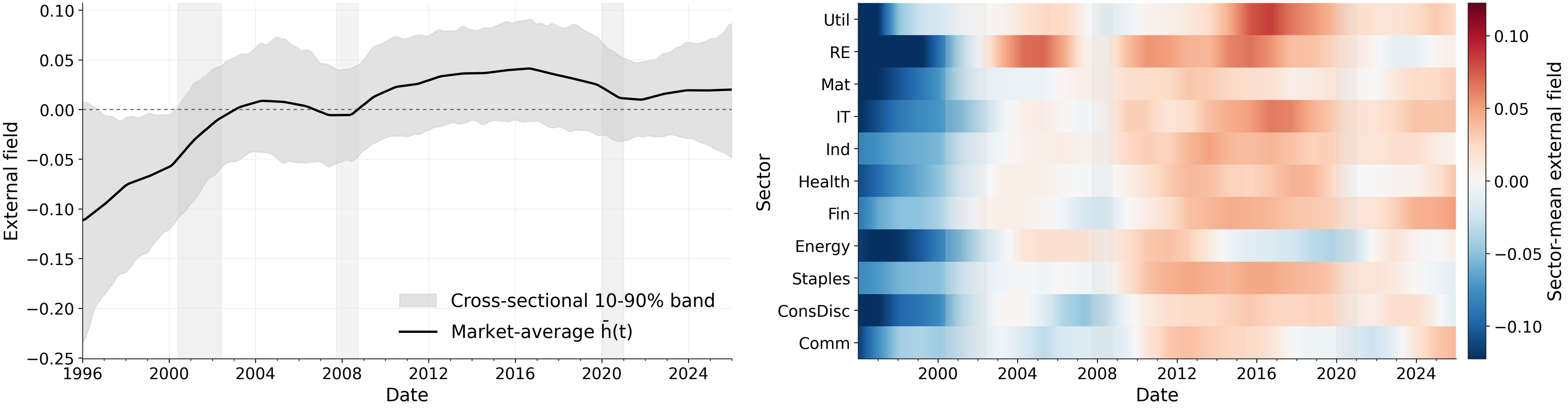}
    \caption{\textbf{External-field dynamics in the kinetic Ising model.} Left: market-average external field $\bar h(t)$ over the full sample; shaded vertical bands mark major market-wide perturbations---the dot-com bust (June 1, 2000--June 1, 2002), the global financial crisis (October 1, 2007--October 1, 2008), and the COVID-19 episode (January 1, 2020--January 1, 2021)---and the gray envelope indicates cross-sectional dispersion across stocks. Right: sector-average external fields through time. Major market perturbations coincide with more diffuse sectoral patterns, whereas quieter periods show more coherent sector-level fields that remain predominantly negative or positive. The initially negative values of $\bar h(t)$ align with the negative empirical market breadth in the early sample, as shown in the left panel of Fig.~\ref{fig:kinetic_market_fit} and summarized in Appendix 
    Table~\ref{tab_apdx:kinetic_diagnostics}. Because $\bar h(t)$ is inferred from binarized open-to-close directional movements and regularized to vary smoothly over time, its sign should be interpreted as a low-frequency baseline tendency in the binary movement process after accounting for interaction and memory effects, rather than as the sign of raw closing-price changes. This distinction also helps explain why short-lived magnitude-driven raw-price drawdowns, such as the April 2025 decline visible in Fig.~\ref{fig:sp500_close_with_hist}, may be attenuated rather 
    than appearing as distinct external-field regimes.}
    \label{fig:kinetic_field_regime}
\end{figure}

Figure~\ref{fig:kinetic_field_regime} shows that the market-average field $\bar h(t)=\tfrac{1}{N}\sum_{i=1}^N h_i(t)$ does not simply mirror empirical market returns. Instead, it evolves at a distinctly low frequency: it is predominantly negative early in the sample, becomes less negative around the dot-com and global financial crisis windows, remains mostly positive through much of the post-2009 period, and softens again during the COVID-19 episode. The sector-level heatmap conveys the same regime structure in cross section. These market-wide perturbation periods are accompanied by more diffuse sectoral field patterns, whereas quieter intervals exhibit more coherent sector-level configurations. In this sense, the external field behaves less like a compressed return series than like a latent background state after accounting for interaction and memory effects. The initially negative regime is consistent with the negative early-sample binary market breadth, visible in the left panel of Fig.~\ref{fig:kinetic_market_fit} and summarized in Appendix Table~\ref{tab_apdx:kinetic_diagnostics}. Because $h_i(t)$ is inferred from binarized open-to-close movements rather than raw prices, the sign of $\bar h(t)$ should be interpreted as a low-frequency tendency in the binary movement process, not as the sign of raw closing-price changes.

This interpretation is consistent with the market-level fit in Fig.~\ref{fig:kinetic_market_fit}. The model reproduces the broad evolution of the empirical market mean, with a Spearman rank correlation of $\rho_s=0.56$ between the two series, indicating a moderate rank-based monotonic association, although it does not capture most of the day-to-day noise. The decomposition of the market-average local field in the right panel helps explain this result. The external field captures the slow drift of market conditions, whereas the interaction term contributes most of the shorter-horizon variation. By contrast, the self-memory component remains small throughout the sample, and the estimated $a_i$ are tightly centered near zero. In this sense, the field sets the background regime, while the interaction term drives most of the short-run adjustment. Next-day market movement in the kinetic model therefore appears to be shaped primarily by changing market conditions and cross-stock interactions rather than by simple own-state persistence.

\begin{figure}[H]
    \centering
    \includegraphics[width=1.0\linewidth]{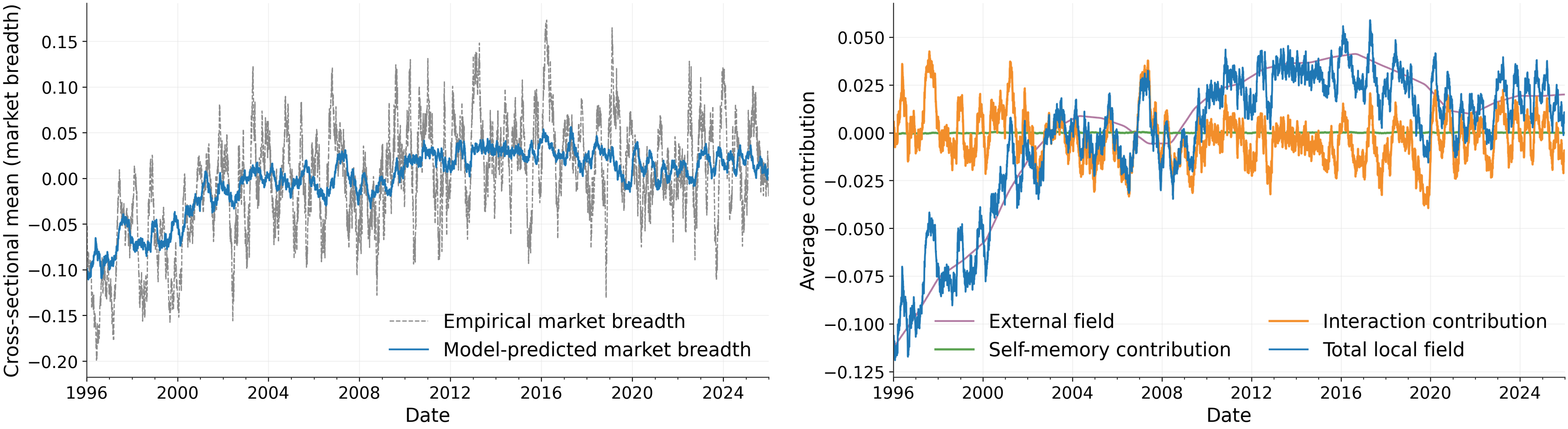}
    \caption{\textbf{Market-level diagnostics for the kinetic model.} Left: model-predicted and empirical market breadths. The model captures the broad evolution of aggregate market movement reasonably well (Spearman $\rho_s=0.56$), although it misses much of the day-to-day variation. Right: decomposition of the market-average local field into external, self-memory, interaction, and total components. The external field mainly tracks the slowly varying market environment, whereas the interaction term contributes most of the higher-frequency variation; the self-memory term remains comparatively small.}
    \label{fig:kinetic_market_fit}
\end{figure}

\begin{figure}[H]
    \centering
    \includegraphics[width=1.0\linewidth]{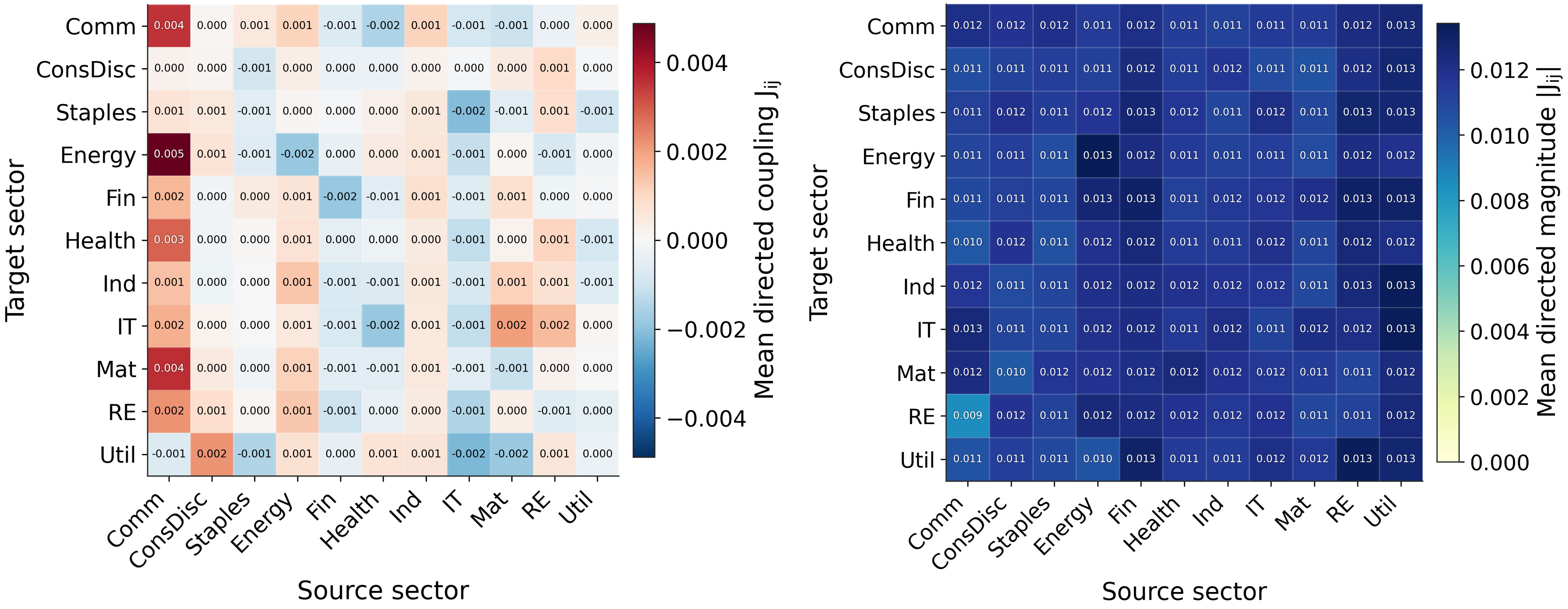}
    \caption{\textbf{Sector-level coupling structure in the kinetic model.} Left: sector-level mean signed directed couplings $J_{ij}$. The matrix is visibly asymmetric, indicating that sector-to-sector influence depends on direction. Right: sector-level mean absolute directed couplings $\lvert J_{ij}\rvert$. Unlike the full-sample static network, the diagonal entries are not strongly dominant, indicating limited additional within-sector concentration in next-day effects.}
    \label{fig:kinetic_sector_couplings}
\end{figure}

Figure~\ref{fig:kinetic_sector_couplings} shows that this interaction layer is organized differently from the long-run static network. In the static model, interaction strength is strongly concentrated within sectors. In the kinetic model, that concentration nearly disappears: the within-sector/between-sector ratio for $\lvert J_{ij}\rvert$ is only 1.016, compared with about 2.8 in the static network, and both the signed and absolute sector interaction is highly asymmetric. Short-horizon propagation is therefore not a directed replica of the static sector blocks. Instead, next-day effects move through a more cross-sector and direction-dependent interaction layer in which source and target are not interchangeable.

\begin{figure}[H]
    \centering
    \includegraphics[width=0.5\linewidth]{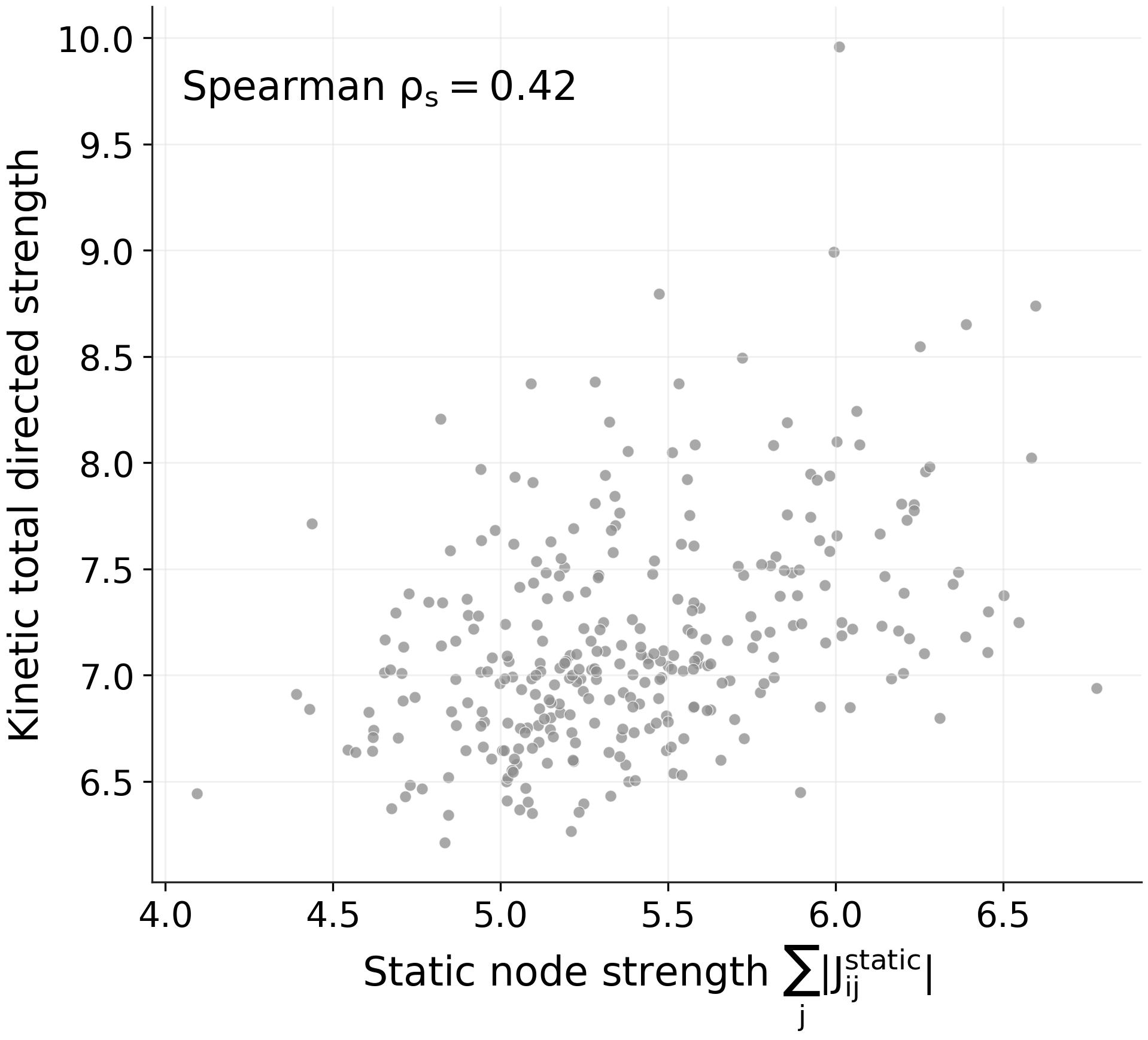}
    \caption{\textbf{Connection between the kinetic and static models.} Static node strength versus kinetic total directed strength. Stocks that are more strongly connected in the full-sample static network also tend to have larger influence in the kinetic model, with a Spearman rank correlation of $\rho_s=0.42$.}
    \label{fig:static_vs_kinetic_strength}
\end{figure}

The comparison with the static network sharpens the central point. Although the parameters of the two models should not be compared one-to-one, stocks that are more strongly embedded in the long-run static network also tend to have larger directed strength in the kinetic model, especially in outgoing influence, as shown in Fig.~\ref{fig:static_vs_kinetic_strength}. The two models are thus aligned in which firms matter most, even though they organize the market in different ways. Overall, the kinetic Ising model portrays the S\&P 500 as a system in which slow regime shifts operate through the external field, while next-day adjustment is governed by a directional and relatively cross-sector transmission network.

\section{Conclusions}\label{sec:conclusion}\noindent
Using a fixed panel of 306 S\&P 500 stocks from 1996 to 2026, this study combined static and kinetic Ising models to characterize the market as both a long-run interaction network and a time-ordered dynamical system. The static model recovered the dominant low-order dependence structure of the panel and revealed a sectorally organized interaction network with modest small-world organization $(\sigma=1.424)$, with within-sector couplings substantially stronger than between-sector couplings. Real estate and energy were especially coherent sectors, and a limited set of firms stood out for simultaneously large field magnitudes and strong aggregate coupling. Although the static fit closely reproduced the empirical first- and second-order moments, the static Ising model is not designed to capture the market’s nonstationary evolution. It is therefore best interpreted as a long-run surrogate for low-order backbone interaction structure rather than as a complete description of the historical sequence.

The kinetic model complemented this picture by incorporating temporal ordering, time-varying external conditions, self-memory, and directed lagged interactions. Its external field captured slow regime shifts, its self-memory terms were generally weak, and its directed couplings formed a much less sector-concentrated transmission layer than the static network. The fitted model was also well calibrated and closely reproduced the one-step statistics it directly constrains. Taken together, the results show that the two models serve distinct but complementary purposes: the static Ising model summarizes persistent interaction structure, whereas the kinetic Ising model captures slowly varying background conditions and next-day cross-stock propagation. More broadly, the combined framework offers a compact way to separate persistent market organization from short-horizon dynamical transmission. Future work could extend this approach through stronger structural regularization for shorter estimation windows, rolling or regime-switching specifications, and higher-order interactions to better represent the full distribution of collective market states.

\bibliography{references}

\section*{Funding Declaration}
\noindent The authors declare no relevant funding.

\section*{Data availability}
\noindent The raw daily open and close price data analyzed in this study were obtained from Yahoo Finance and are publicly available subject to its terms of use. The processed data supporting the findings of this study are available from the corresponding authors upon reasonable request.

\section*{Code availability}
\noindent The code used to generate the results is available at [URL that will be available once the paper is accepted].

\section*{Competing interests}
\noindent The authors declare no competing interests.

\section*{Author contributions}
\noindent S.O. conceived the study, collected and processed the data, developed the models, performed the analyses, prepared the figures, and wrote the original manuscript. Z.W. and M.C.G. provided supervision and guidance, contributed to the interpretation of the results, and reviewed and edited the manuscript. All authors reviewed and approved the final manuscript.

\appendix
\setcounter{figure}{0}
\setcounter{table}{0}
\renewcommand{\thefigure}{A.\arabic{figure}}
\renewcommand{\thetable}{A.\arabic{table}}

\section{Maximum entropy derivation of the static and kinetic Ising models}\label{app:maxent}\noindent
This appendix outlines the maximum entropy derivations of the static and kinetic Ising models used in the main text. In each case, the model is obtained by maximizing entropy subject to normalization and the relevant moment constraints, with the resulting coefficients given by the associated Lagrange multipliers.

\subsection{Static pairwise Ising model}\noindent
Let $\Omega=\{-1,+1\}^N$ denote the set of spin configurations, indexed by $\vect{s}^{(k)}$, $k=1,\ldots,2^N$, and write $P_k=\Prob{\{\vect{s}^{(k)}\}}$. We maximize
\begin{equation}
    \mathcal{S}=-\sum_{k=1}^{2^N} P_k \log P_k
\end{equation}
subject to
\begin{equation}
    \sum_{k=1}^{2^N} P_k = 1,
\end{equation}
\begin{equation}
    \sum_{k=1}^{2^N} P_k s_i^{(k)} = m_i,
    \qquad
    \sum_{k=1}^{2^N} P_k s_i^{(k)} s_j^{(k)} = C_{ij},
    \quad i<j.
\end{equation}
Introducing multipliers $\alpha$, $h_i$, and $J_{ij}$ for these constraints gives
\begin{equation}
\begin{aligned}
    \mathcal{L}
    =&
    -\sum_{k=1}^{2^N} P_k \log P_k
    -\alpha\left(1-\sum_{k=1}^{2^N} P_k\right) \\
    &-\sum_{i=1}^N h_i\left(m_i-\sum_{k=1}^{2^N} P_k s_i^{(k)}\right)
    -\sum_{i<j} J_{ij}\left(C_{ij}-\sum_{k=1}^{2^N} P_k s_i^{(k)} s_j^{(k)}\right).
\end{aligned}
\end{equation}
Setting $\partial \mathcal{L}/\partial P_k=0$ yields
\begin{equation}
    -\log P_k - 1 + \alpha
    + \sum_{i=1}^N h_i s_i^{(k)}
    + \sum_{i<j} J_{ij} s_i^{(k)} s_j^{(k)}
    =0,
\end{equation}
so
\begin{equation}
    P_k
    =
    \frac{1}{Z}
    \exp\!\left(
        \sum_{i=1}^N h_i s_i^{(k)}
        + \sum_{i<j} J_{ij} s_i^{(k)} s_j^{(k)}
    \right),
\end{equation}
with partition function
\begin{equation}
    Z
    =
    \sum_{k=1}^{2^N}
    \exp\!\left(
        \sum_{i=1}^N h_i s_i^{(k)}
        + \sum_{i<j} J_{ij} s_i^{(k)} s_j^{(k)}
    \right).
\end{equation}
Equivalently, for a generic configuration $\vect{s}$,
\begin{equation}
    p(\vect{s})
    =
    \frac{1}{Z}
    \exp\!\left(
        \sum_{i=1}^N h_i s_i
        + \sum_{i<j} J_{ij} s_i s_j
    \right),
\end{equation}
which is the static pairwise Ising model.

\subsection{Kinetic Ising model}\noindent
The kinetic model follows from an analogous conditional maximum entropy construction. For each current state $\vect{s}(t)$, let $p_t(\vect{s}'\mid\vect{s}(t))$ denote the one-step transition law to a generic next state $\vect{s}'$. We maximize the conditional entropy
\begin{equation}
    \mathcal{S}_{\mathrm{c}}
    =
    -\sum_{t=1}^{T-1}
    \sum_{\vect{s}'\in\{-1,+1\}^N}
    p_t(\vect{s}'\mid\vect{s}(t))
    \log p_t(\vect{s}'\mid\vect{s}(t))
\end{equation}
subject to normalization at each time point,
\begin{equation}
    \sum_{\vect{s}'\in\{-1,+1\}^N}
    p_t(\vect{s}'\mid\vect{s}(t))
    =1,
    \qquad t=1,\ldots,T-1,
\end{equation}
and constraints on the basis-weighted next-step means, self-lag products, and cross-lag products:
\begin{equation}
    \sum_{t=1}^{T-1}
    \sum_{\vect{s}'}
    p_t(\vect{s}'\mid\vect{s}(t))\,\phi_m(t)\,s_i'
    =
    B_{im},
\end{equation}
\begin{equation}
    \sum_{t=1}^{T-1}
    \sum_{\vect{s}'}
    p_t(\vect{s}'\mid\vect{s}(t))\,s_i' s_i(t)
    =
    D_i,
\end{equation}
\begin{equation}
    \sum_{t=1}^{T-1}
    \sum_{\vect{s}'}
    p_t(\vect{s}'\mid\vect{s}(t))\,s_i' s_j(t)
    =
    K_{ij},
    \qquad j\neq i,
\end{equation}
where $B_{im}$, $D_i$, and $K_{ij}$ are the corresponding empirical quantities.

Introducing multipliers $\lambda_t$, $\gamma_{im}$, $a_i$, and $J_{ij}$ gives
\begin{equation}
\begin{aligned}
    \mathcal{L}_{\mathrm{c}}
    =&
    -\sum_{t=1}^{T-1}
    \sum_{\vect{s}'}
    p_t(\vect{s}'\mid\vect{s}(t))
    \log p_t(\vect{s}'\mid\vect{s}(t)) \\
    &-\sum_{t=1}^{T-1}
    \lambda_t
    \left(
        1-\sum_{\vect{s}'} p_t(\vect{s}'\mid\vect{s}(t))
    \right) \\
    &-\sum_{i=1}^N\sum_{m=1}^{M}
    \gamma_{im}
    \left(
        B_{im}
        -\sum_{t=1}^{T-1}\sum_{\vect{s}'} p_t(\vect{s}'\mid\vect{s}(t))\,\phi_m(t)\,s_i'
    \right) \\
    &-\sum_{i=1}^N
    a_i
    \left(
        D_i
        -\sum_{t=1}^{T-1}\sum_{\vect{s}'} p_t(\vect{s}'\mid\vect{s}(t))\,s_i' s_i(t)
    \right) \\
    &-\sum_{i=1}^N\sum_{j\neq i}
    J_{ij}
    \left(
        K_{ij}
        -\sum_{t=1}^{T-1}\sum_{\vect{s}'} p_t(\vect{s}'\mid\vect{s}(t))\,s_i' s_j(t)
    \right).
\end{aligned}
\end{equation}
Setting the derivative with respect to $p_t(\vect{s}'\mid\vect{s}(t))$ to zero gives
\begin{equation}
\begin{aligned}
    -\log p_t(\vect{s}'\mid\vect{s}(t))
    -1
    +\lambda_t
    &+\sum_{i=1}^N\sum_{m=1}^{M}\gamma_{im}\phi_m(t)s_i' \\
    &+\sum_{i=1}^N a_i s_i' s_i(t)
    +\sum_{i=1}^N\sum_{j\neq i} J_{ij}s_i' s_j(t)
    =0.
\end{aligned}
\end{equation}
Hence
\begin{equation}
    p_t(\vect{s}'\mid\vect{s}(t))
    =
    \frac{1}{Z_t(\vect{s}(t))}
    \exp\!\left[
    \sum_{i=1}^N s_i'
    \left(
    \sum_{m=1}^{M}\phi_m(t)\gamma_{im}
    +a_i s_i(t)
    +\sum_{j\neq i} J_{ij}s_j(t)
    \right)
    \right].
\end{equation}
Defining the effective local field as
\begin{equation}
    \theta_i(t)
    =
    \sum_{m=1}^{M}\phi_m(t)\gamma_{im}
    +a_i s_i(t)
    +\sum_{j\neq i} J_{ij}s_j(t),
\end{equation}
we obtain
\begin{equation}
    p_t(\vect{s}'\mid\vect{s}(t))
    =
    \frac{1}{Z_t(\vect{s}(t))}
    \exp\!\left(
    \sum_{i=1}^N s_i' \theta_i(t)
    \right).
\end{equation}
Because the exponent separates over $i$, the normalizing constant factorizes as
\begin{equation}
    Z_t(\vect{s}(t))
    =
    \prod_{i=1}^{N} 2\cosh\!\left(\theta_i(t)\right),
\end{equation}
and therefore
\begin{equation}
    p_t(\vect{s}'\mid\vect{s}(t))
    =
    \prod_{i=1}^{N}
    \frac{\exp\!\left(s_i' \theta_i(t)\right)}
    {2\cosh\!\left(\theta_i(t)\right)}.
\end{equation}
Replacing $\vect{s}'$ with the empirical next state $\vect{s}(t+1)$ gives
\begin{equation}
    p\!\left(\vect{s}(t+1)\mid\vect{s}(t)\right)
    =
    \prod_{i=1}^{N}
    \frac{\exp\!\left(s_i(t+1)\theta_i(t)\right)}
    {2\cosh\!\left(\theta_i(t)\right)},
\end{equation}
which is the synchronous kinetic Ising model.

\newpage
\begin{figure}[H]
    \centering
    \includegraphics[width=1.0\linewidth]{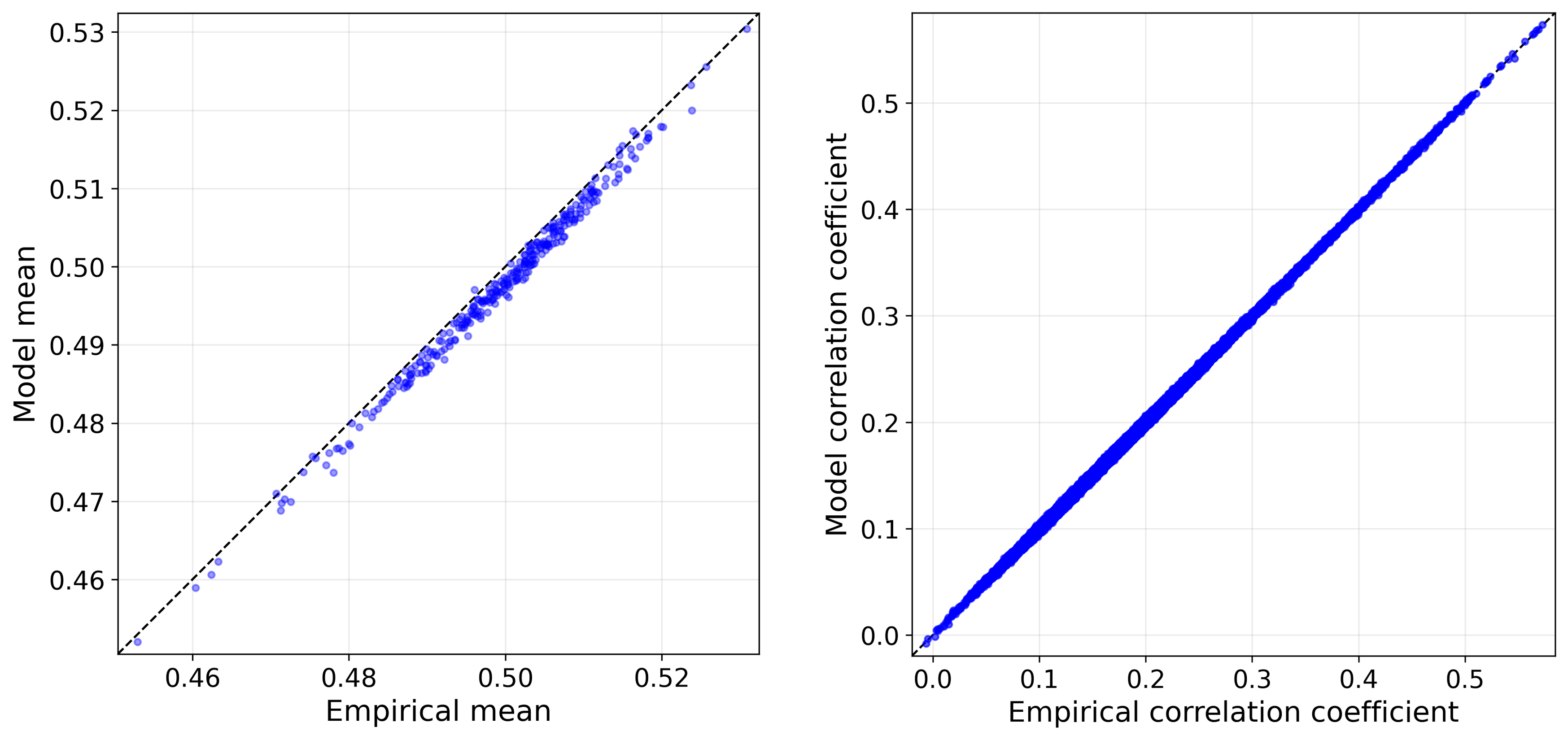}
    \caption{\textbf{Validation of the fitted static Ising model against the empirical constraints.} Left: empirical mean $\mathcal{M}_{1,i}^\mathcal{D}$ versus model mean $\mathcal{M}_{1,i}$ across stocks. Right: empirical versus model pairwise correlation coefficient, excluding the trivial diagonal entries equal to 1. The dashed line is the identity line. The near-perfect alignment confirms that the fitted model closely reproduces the first- and second-order moments used in estimation.}
    \label{fig_apdx:validation_static}
\end{figure}

\newpage
\begin{figure}[H]
    \centering
    \includegraphics[width=1.0\linewidth]{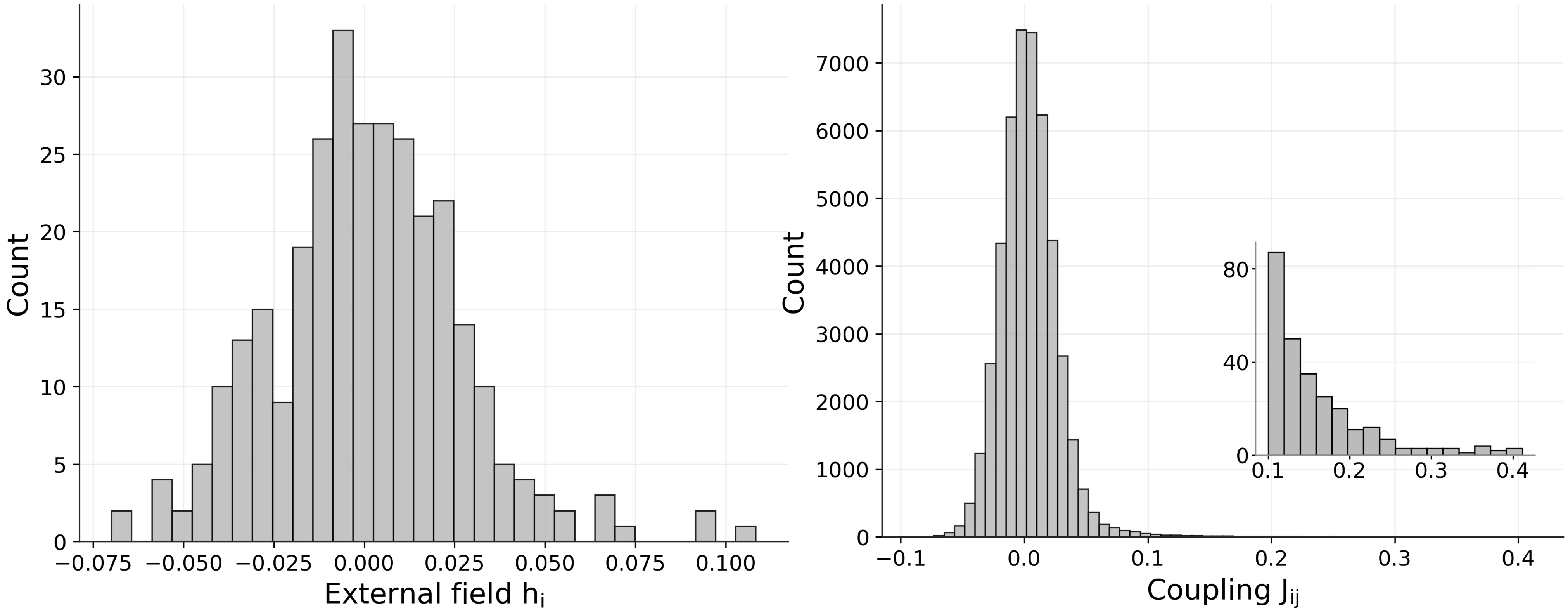}
    \caption{\textbf{Distributions of the inferred external fields and couplings in the static Ising model.} Left: histogram of the estimated external fields $h_i$. Right: histogram of the estimated pairwise couplings $J_{ij}$. The $h_i$ values are centered near zero with moderate spread, whereas the $J_{ij}$ values are sharply concentrated near zero, indicating that most inferred pairwise interactions are weak. The inset highlights the positive tail of the coupling distribution.}
    \label{fig_apdx:static_ising_params_histograms}
\end{figure}

\newpage
\begin{table}[H]
    \centering
    \caption{\textbf{Detailed diagnostics of the static Ising interaction network for the full 30-year S\&P 500 sample.}\tnote{*}\tnote{$\dagger$}}
    \label{tab_apdx:static_network_diagnostics}
    \small
    \renewcommand{\arraystretch}{1.08}
    \begin{threeparttable}
    \begin{tabularx}{\textwidth}{@{}p{0.67\textwidth}>{\raggedleft\arraybackslash}X@{}}
    \toprule
    Description & Value \\
    \midrule
    
    \multicolumn{2}{@{}l}{\textbf{Filtering and graph size}} \\
    Number of stocks / unique pairwise couplings & 306 / 46{,}665 \\
    Filtering rule & Top decile of $\lvert J_{ij}\rvert$ \\
    Top-decile cutoff on $\lvert J_{ij}\rvert$ & 0.03534 \\
    Retained edges / edge fraction & 4{,}667 / 0.1000 \\
    Connected components & 1 \\
    
    \addlinespace[4pt]
    \multicolumn{2}{@{}l}{\textbf{Filtered-network diagnostics}} \\
    Nodes / edges & 306 / 4{,}667 \\    
    Average degree & 30.50 \\
    Average clustering coefficient & 0.1450 \\
    Average shortest-path length (largest connected component) & 1.9670 \\
    Number of positive / negative edges & 3{,}317 / 1{,}350 \\
    Sector assortativity & 0.259 \\
    Small-world coefficient $\sigma$ & 1.424 \\
    
    \addlinespace[4pt]
    \multicolumn{2}{@{}l}{\textbf{Sector-level structure}} \\
    Within-sector mean $\lvert J_{ij}\rvert$ / between-sector mean $\lvert J_{ij}\rvert$ & 0.04367 / 0.01566 \\
    Within/between ratio & 2.7879 \\

    \multirow[c]{5}{=}{Largest within-sector mean $\lvert J_{ij}\rvert$} 
        & \multicolumn{1}{@{}r@{}}{Real Estate (0.0737)} \\
        & \multicolumn{1}{@{}r@{}}{Energy (0.0710)} \\
        & \multicolumn{1}{@{}r@{}}{Utilities (0.0526)} \\
        & \multicolumn{1}{@{}r@{}}{Communication Services (0.0525)} \\
        & \multicolumn{1}{@{}r@{}}{Materials (0.0412)} \\

    \addlinespace[5pt]

    \multirow[c]{5}{=}{Largest sector mean $\lvert h_i\rvert$} 
        & \multicolumn{1}{@{}r@{}}{Information Technology (0.0259)} \\
        & \multicolumn{1}{@{}r@{}}{Financials (0.0210)} \\
        & \multicolumn{1}{@{}r@{}}{Consumer Discretionary (0.0209)} \\
        & \multicolumn{1}{@{}r@{}}{Consumer Staples (0.0198)} \\
        & \multicolumn{1}{@{}r@{}}{Materials (0.0198)} \\
    
    \addlinespace[4pt]
    \multicolumn{2}{@{}l}{\textbf{Stock-level diagnostics}} \\
    
    \multirow[c]{5}{=}{Largest $\lvert h_i\rvert$} 
        & \multicolumn{1}{@{}r@{}}{Ford (0.1083)} \\
        & \multicolumn{1}{@{}r@{}}{AMD (0.0933)} \\
        & \multicolumn{1}{@{}r@{}}{ODFL (0.0923)} \\
        & \multicolumn{1}{@{}r@{}}{CRH (0.0712)} \\
        & \multicolumn{1}{@{}r@{}}{PG (0.0699)} \\
    
    \addlinespace[5pt]
    
    \multirow[c]{5}{=}{Largest $\sum_j \lvert J_{ij}\rvert$} 
        & \multicolumn{1}{@{}r@{}}{WEC (3.047)} \\
        & \multicolumn{1}{@{}r@{}}{BAC (2.991)} \\
        & \multicolumn{1}{@{}r@{}}{XOM (2.938)} \\
        & \multicolumn{1}{@{}r@{}}{AVB (2.906)} \\
        & \multicolumn{1}{@{}r@{}}{AMAT (2.885)} \\
    
    \bottomrule
    \end{tabularx}
    
    \begin{tablenotes}[flushleft]
    \footnotesize
    \item[*] Abbreviations: AMD = Advanced Micro Devices; ODFL = Old Dominion Freight Line; CRH = CRH plc; PG = Procter \& Gamble; WEC = WEC Energy Group; BAC = Bank of America; XOM = ExxonMobil; AVB = AvalonBay Communities; AMAT = Applied Materials.
    \item[$\dagger$] The cutoff on $\lvert J_{ij} \rvert$ is obtained by sorting the $306\times305/2=46{,}665$ unique pairwise coupling magnitudes $\lvert J_{ij} \rvert$ in decreasing order. The top-decile threshold is then used as the cutoff, so retaining edges with $\lvert J_{ij} \rvert$ above this value yields $4{,}667$ edges in the filtered network.
    \end{tablenotes}
    \end{threeparttable}
\end{table}

\newpage
\begin{figure}[H]
    \centering
    \includegraphics[width=1.0\linewidth]{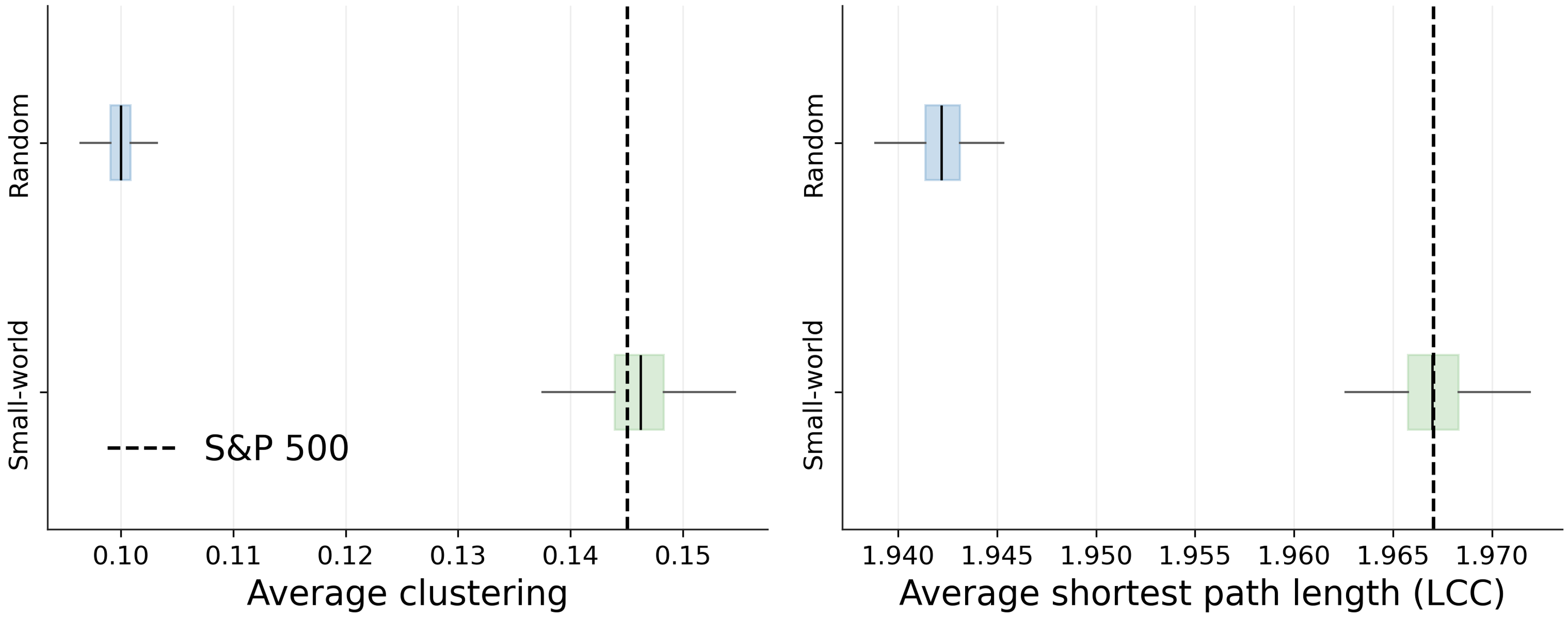}
    \caption{\textbf{Benchmark comparison for the filtered stock interaction network.} Left: average clustering coefficient. Right: average shortest-path length. Boxplots show the distributions of these statistics across 300 realizations each of edge-matched random graphs and Watts--Strogatz graphs~\cite{watts1998collective} with comparable mean degree, with the latter generated using a rewiring probability of $\beta=0.53$. In each panel, the dashed vertical line marks the observed value for the filtered S\&P 500 network. The observed network exhibits substantially higher clustering than the random benchmark while maintaining a similarly short path length, placing it closer to the small-world benchmark in clustering and to the random benchmark in path length, consistent with small-world organization.}
    \label{fig_apdx:benchmark_comparison}
\end{figure}

\newpage
\begin{figure}[H]
    \centering
    \includegraphics[width=0.6\linewidth]{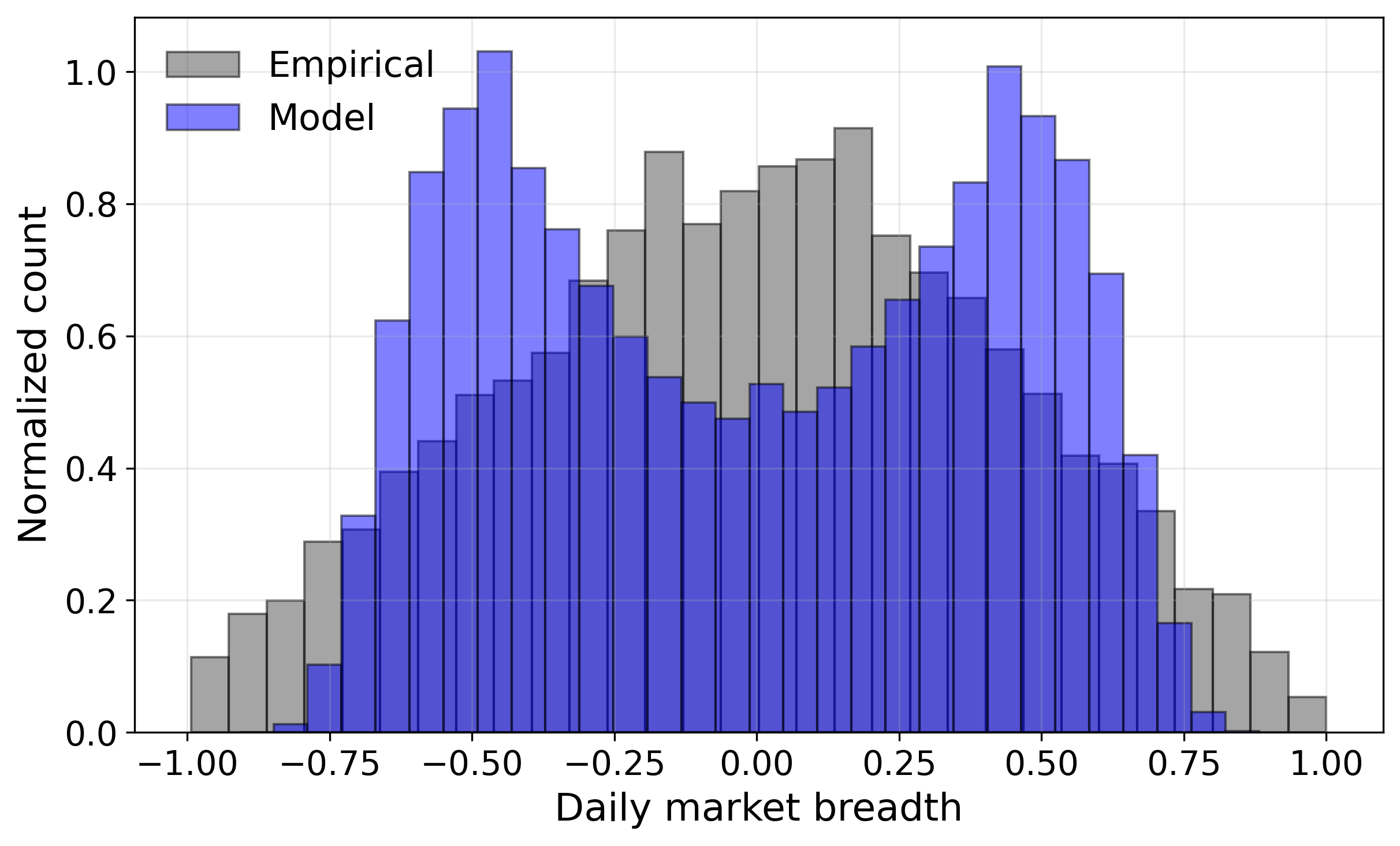}
    \caption{\textbf{Distribution of daily market breadth in the empirical data and in equilibrium samples from the fitted static Ising model.} Here the daily market breadth is defined as $m=\tfrac{1}{N}\sum_{i=1}^{N}s_i$. The empirical distribution is unimodal, whereas the model distribution is more clearly bimodal, showing that the static fit reproduces low-order moments well but does not capture the full distribution of collective market states.}
    \label{fig_apdx:validation_static_magnetization}
\end{figure}

\newpage
\begin{figure}[H]
    \centering
    \includegraphics[width=1.0\linewidth]{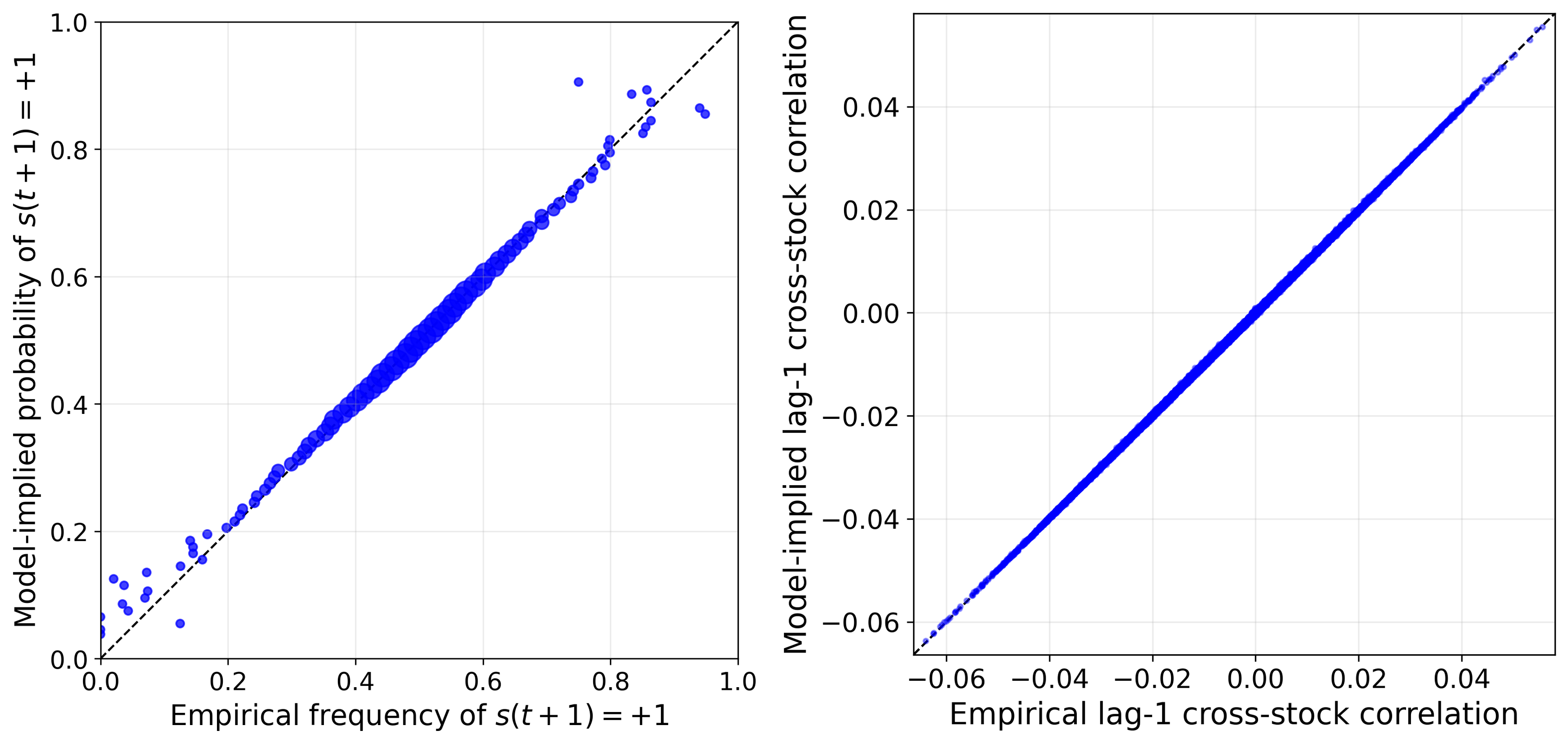}
    \caption{\textbf{Validation of the fitted kinetic Ising model.} Left: comparison between the empirical frequency of an upward next-day move and the corresponding model-implied probability, $P(s_i(t+1)=+1|\vect{s}(t))$, grouped into 100 bins (equivalently, 0.01 probability increment) to assess calibration. This quantity measures how well the fitted kinetic Ising model reproduces the observed frequencies of $s(t+1)=+1$. Each point represents one non-empty bin of stock-day observations, and marker size is proportional to the number of observations in the bin. Right: comparison between empirical and model-implied off-diagonal lag-1 cross-stock correlations, $\mathrm{corr}\!\left(s_i(t+1),\,s_j(t)\right)$ where $i\neq j$, which measure how the state of stock $j$ at day $t$ is associated with the next-day state of stock $i$. Each point in the right panel represents one ordered stock pair $(i,j)$ with $i\neq j$. The dashed line in each panel is the identity line; closer agreement indicates more accurate reproduction of the corresponding one-step statistic by the fitted kinetic Ising model.}
    \label{fig_apdx:validation_kinetic}
\end{figure}

\newpage
\begin{table}[H]
    \centering
    \caption{\textbf{Detailed diagnostics of the kinetic Ising model for the full 30-year S\&P 500 sample.}\tnote{*}\tnote{$\dagger$}}
    \label{tab_apdx:kinetic_diagnostics}
    \small
    \renewcommand{\arraystretch}{1.08}
    \begin{threeparttable}
    \begin{tabularx}{\textwidth}{@{}p{0.67\textwidth}>{\raggedleft\arraybackslash}X@{}}
    \toprule
    Description & Value \\
    \midrule
    
    \multicolumn{2}{@{}l}{\textbf{Sample and model size}} \\
    Number of stocks / one-step transitions / basis count & 306 / 7{,}549 / 30 \\

    \addlinespace[4pt]
    \multicolumn{2}{@{}l}{\textbf{Market-level fit and regime diagnostics}} \\
    Spearman $\rho_s$ between model-predicted and empirical market means & 0.56 \\
    Full-sample mean $\bar h(t)$ / standard deviation $\bar h(t)$ & 0.00157 / 0.03812 \\
    Spearman $\rho_s\!\left(\bar h(t), \text{empirical market mean}\right)$ & 0.09095 \\
    Spearman $\rho_s\!\left(\bar\theta(t), \text{empirical market mean}\right)$ & 0.21719 \\

    \addlinespace[4pt]
    \multicolumn{2}{@{}l}{\textbf{Windowed field diagnostics}} \\
    \multicolumn{2}{@{}p{\dimexpr\linewidth-2\tabcolsep\relax}@{}}%
    {\footnotesize Values in parentheses are mean $\bar h(t)$, mean $\bar\theta(t)$, empirical market mean, respectively.} \\
    Full sample  & $\phantom{-}0.00157,\,-0.00132,\,-0.00115$ \\
    Early sample & $-0.04353,\,-0.04217,\,-0.04473$ \\
    Dot-com bust & $-0.02604,\,-0.02302,\,-0.02085$ \\
    GFC          & $-0.00185,\,-0.01180,\,-0.04717$ \\
    COVID        & $\phantom{-}0.01298,\;\phantom{-}0.01152,\;\phantom{-}0.00098$ \\
    
    \addlinespace[4pt]
    \multicolumn{2}{@{}l}{\textbf{Self-memory terms}} \\
    Mean $a_i$ / median $a_i$ & 0.00119 / 0.00033 \\
    Fraction positive / negative $a_i$ & 0.50654 / 0.49346 \\

    \multirow[c]{5}{=}{Largest positive $a_i$}
        & \multicolumn{1}{@{}r@{}}{AIG (0.0626)} \\
        & \multicolumn{1}{@{}r@{}}{CRH (0.0549)} \\
        & \multicolumn{1}{@{}r@{}}{TPL (0.0473)} \\
        & \multicolumn{1}{@{}r@{}}{APA (0.0410)} \\
        & \multicolumn{1}{@{}r@{}}{EME (0.0395)} \\

    \addlinespace[5pt]

    \multirow[c]{5}{=}{Most negative $a_i$}
        & \multicolumn{1}{@{}r@{}}{NEM ($-0.0568$)} \\
        & \multicolumn{1}{@{}r@{}}{QCOM ($-0.0498$)} \\
        & \multicolumn{1}{@{}r@{}}{EMR ($-0.0443$)} \\
        & \multicolumn{1}{@{}r@{}}{ECL ($-0.0391$)} \\
        & \multicolumn{1}{@{}r@{}}{MTB ($-0.0371$)} \\
    
    \addlinespace[4pt]
    \multicolumn{2}{@{}l}{\textbf{Directed-coupling diagnostics}} \\
    Mean $J_{ij}$ / standard deviation $J_{ij}$ & $-0.00010$ / 0.01474 \\
    Mean $\lvert J_{ij}\rvert$ / 90th percentile $\lvert J_{ij}\rvert$ & 0.01174 / 0.02425 \\
    Within-sector mean $\lvert J_{ij}\rvert$ / between-sector mean $\lvert J_{ij}\rvert$ & 0.01185 / 0.01167 \\
    Within/between ratio & 1.0161 \\
    Frobenius asymmetry index & 1.415 \\
    Symmetry Spearman $\rho_s(J_{ij},J_{ji})$ / Pearson $r(J_{ij},J_{ji})$ & $-0.0013$ / $-0.0017$ \\

    \addlinespace[4pt]
    \multicolumn{2}{@{}l}{\textbf{Comparison with the static network}} \\
    Static within-sector mean $\lvert J_{ij}\rvert$ / between-sector mean $\lvert J_{ij}\rvert$ & 0.04367 / 0.01566 \\
    Static within/between ratio & 2.7879 \\
    Spearman $\rho_s$ (static strength, kinetic total / incoming / outgoing strength) & 0.41558 / $-0.01553$ / 0.46854 \\

    \bottomrule
    \end{tabularx}

    \begin{tablenotes}[flushleft]
    \footnotesize
    \item[*] Here $\bar h(t)=\tfrac{1}{N}\sum_{i=1}^N h_i(t)$ and $\bar\theta(t)=\tfrac{1}{N}\sum_{i=1}^N \theta_i(t)$ denote the market-average external field and market-average total local field, respectively; the empirical market mean is $\tfrac{1}{N}\sum_{i=1}^N s_i(t)$. All reported $\rho_s$ values are Spearman rank correlations. Window definitions: Early sample = 1996-01-01 to 2004-12-31; Dot-com bust = 2000-06-01 to 2002-06-01; GFC = 2007-10-01 to 2008-10-01; COVID = 2020-01-01 to 2021-01-01.
    \item[$\dagger$] Abbreviations: AIG = American International Group; CRH = CRH plc; TPL = Texas Pacific Land Corporation; APA = APA Corporation; EME = Emcor; NEM = Newmont; QCOM = Qualcomm; EMR = Emerson Electric; ECL = Ecolab; MTB = M\&T Bank; GFC = global financial crisis.
    \end{tablenotes}
    \end{threeparttable}
\end{table}

\end{document}